\documentclass[lettersize,journal]{IEEEtran}

\usepackage{amsmath,amsfonts}
\usepackage{algorithmic}
\usepackage{algorithm}
\usepackage{array}
\usepackage[caption=false,font=normalsize,labelfont=sf,textfont=sf]{subfig}
\usepackage{textcomp}
\usepackage{stfloats}
\usepackage{url}
\usepackage{verbatim}
\usepackage{graphicx}
\usepackage{cite}

\usepackage{booktabs}
\usepackage{makecell}
\usepackage{multirow}
\usepackage{multicol}
\usepackage{tabularx}
\usepackage[dvipsnames]{xcolor}
\usepackage{caption}
\usepackage{shellesc}
\usepackage{tablefootnote}

\usepackage{nomencl}
\makenomenclature

\usepackage{glossaries-extra}

\usepackage{glossaries-extra}
\setabbreviationstyle[acronym]{long-short}

\newacronym{ss}{speaker separation}{speaker separation}
\newacronym{se}{speech enhancement}{speech enhancement}
\newacronym{tse}{TSE}{target speaker extraction}
\newacronym{sex}{SEx}{speaker extraction}

\newacronym{ao}{AO}{audio-only}
\newacronym{av}{AV}{audiovisual}
\newacronym{avse}{AVSE}{audiovisual speech enhancement}
\newacronym{avss}{AVSS}{audiovisual speaker separation}
\newacronym{avsex}{AVSEx}{audiovisual speaker extraction}

\newacronym{istft}{iSTFT}{inverse short-time Fourier transform}
\newacronym{stft}{STFT}{short-time Fourier transform}
\newacronym{tf}{T-F}{time-frequency}
\newacronym{rir}{RIR}{room impulse responses}
\newacronym{mcmfwf}{MCMFWF}{multi-channel multi-frame Wiener filter}
\newacronym{scmfwf}{SCMFWF}{single-channel multi-frame Wiener filter}
\newacronym{asr}{ASR}{automatic speech recognition}
\newacronym{ri}{RI}{real and imaginary}

\newacronym{ffn}{FFN}{feed-forward networks}
\newacronym{deconv1d}{Deconv1D}{one-dimensional deconvolution}
\newacronym{mhsa}{MHA}{multi-head self-attention}
\newacronym{gmhsa}{GMHSA}{global multi-head self-attention}
\newacronym{ln}{LN}{layer normalization}
\newacronym{relu}{ReLU}{rectified linear unit}
\newacronym{prelu}{PReLU}{parametric rectified linear unit}
\newacronym{lnorm}{LN}{layer normalization}
\newacronym{glnorm}{gLN}{global layer normalization}
\newacronym{gnorm}{GN}{group normalization}
\newacronym{conv2d}{Conv2D}{2D convolutional layer}
\newacronym{conv1d}{Conv1D}{1D convolutional layer}
\newacronym{dnn}{DNN}{deep neural network}
\newacronym{cnn}{CNN}{convolutional neural network}
\newacronym{rnn}{RNN}{recurrent neural network}
\newacronym{lstm}{LSTM}{long short-term memory}
\newacronym{blstm}{BLSTM}{bidirectional long short-term memory}
\newacronym{gru}{GRU}{gated recurrent unit}
\newacronym{sgd}{SGD}{stochastic gradient descent}
\newacronym{adam}{Adam}{adaptive moment estimation}
\newacronym{bptt}{BPTT}{backpropagation through time}
\newacronym{nlp}{NLP}{natural language processing}
\newacronym{cv}{CV}{computer vision}
\newacronym{dm}{DM}{dynamic mixing}
\newacronym{pe}{PE}{positional encoding}
\newacronym{rpe}{RPE}{relative positional encoding}
\newacronym{rcpe}{RCPE}{random-chunk positional encoding}
\newacronym{tconvffn}{T-ConvFFN}{time-convolutional feed-forward network}
\newacronym{tconv}{T-Conv}{time-convolutional}
\newacronym{tgcon1d}{T-GConv1D}{grouped 1d convolution}

\newacronym{silu}{SiLU}{sigmoid-weighted linear unit}
\newacronym{pit}{PIT}{permutation invariant training}
\newacronym{nb-mhsa}{NB-MHSA}{narrow-band multi-head self-attention}

\newacronym{pesq}{PESQ}{perceptual evaluation of speech quality}
\newacronym{nb-pesq}{PESQ}{narrow-band perceptual evaluation of speech quality}
\newacronym{snr}{SNR}{signal-to-noise ratio}
\newacronym{sdr}{SDR}{signal-to-distortion ratio}
\newacronym{sdri}{SDRi}{signal-to-distortion ratio improvement}
\newacronym{si-sdr}{SI-SDR}{scale-invariant signal-to-distortion ratio}
\newacronym{si-sdrse}{SI-SDR(SE)}{scale-invariant signal-to-distortion ratio (scaled estimated)}
\newacronym{si-sdri}{SI-SDRi}{scale-invariant signal-to-distortion ratio improvement}
\newacronym{stoi}{STOI}{short-time objective intelligibility} 
\newacronym{estoi}{eSTOI}{extended short-time objective intelligibility} 
\newacronym{mc}{MC}{mixture constraint loss}
\newacronym{wer}{WER}{word error rate}

\newacronym{fps}{FPS}{frames per second}

\newacronym{flop}{FLOP}{floating point operation}
\newacronym{gflop}{GFLOP}{Giga floating point operation}

\usepackage{pifont}
\newcommand{\cmark}{\ding{51}}%
\newcommand{\xmark}{\ding{55}}%

\usepackage[hidelinks]{hyperref} %

\makenomenclature

\begin{document}

\title{CrossNet: Leveraging Global, Cross-Band, Narrow-Band, and Positional Encoding for Single- and Multi-Channel Speaker Separation}

\author{
Vahid Ahmadi Kalkhorani,
DeLiang Wang, ~\IEEEmembership{Fellow,~IEEE}
\thanks{
Vahid Ahmadi Kalkhorani is with the Department of Computer Science and Engineering, The Ohio State University, Columbus, OH 43210 USA (e-mail: ahmadikalkhorani.1@osu.edu).
DeLiang Wang is with the Department of Computer Science and Engineering, and the
Center for Cognitive and Brain Sciences, The Ohio State University, Columbus,
OH 43210 USA (e-mail: dwang@cse.ohio-state.edu).
}
}



\maketitle

\begin{abstract}
    We introduce CrossNet, a complex spectral mapping approach to speaker separation and enhancement in reverberant and noisy conditions. The proposed architecture comprises an encoder layer, a global multi-head self-attention module, a cross-band module, a narrow-band module, and an output layer. CrossNet captures global, cross-band, and narrow-band correlations in the time-frequency domain. To address performance degradation in long utterances, we introduce a random chunk positional encoding. Experimental results on multiple datasets demonstrate the effectiveness and robustness of CrossNet, achieving state-of-the-art performance in tasks including reverberant and noisy-reverberant speaker separation. Furthermore, CrossNet exhibits faster and more stable training in comparison to recent baselines. Additionally, CrossNet's high performance extends to multi-microphone conditions, demonstrating its versatility in various acoustic scenarios.
\end{abstract}


\section{Introduction}

\IEEEPARstart{I}{n} human and machine speech communication, the presence of acoustic interference, such as background noise or competing speakers, presents a considerable challenge for speech understanding. To address these challenges, speech separation systems have been developed to separate target speech signals from noisy and reverberant environments. Speech separation includes speaker separation and speech enhancement \cite{wang2018supervised}. The task of speaker separation is to separate the speech signals of multiple speakers and speech enhancement aims to separate a single speech signal from nonspeech background noise. Both tasks are essential for various applications, including hearing aids, teleconferencing, and voice-controlled assistants.

Significant strides have been made in monaural talker-independent speaker separation with the introduction of deep clustering \cite{hershey2016deep} and \gls{pit} \cite{kolbaek2017multitalker}. By effectively tackling the permutation ambiguity issue inherent in talker-independent training, these approaches have substantially elevated speaker separation performance. Subsequent developments have produced impressive performance gains. 

For example, deep CASA \cite{liu2019divide} breaks down the speaker separation task into two phases: simultaneous grouping and sequential grouping. 
Conv-TasNet \cite{luo2019conv} operates on short windows of signals and performs end-to-end masking-based separation. 
DPRNN \cite{luo2020dprnn} segments a time-domain signal into fixed-length blocks, where intra- and inter-block \glspl{rnn} are applied iteratively to facilitate both local and global processing. 
SepFormer \cite{subakan2021attention} replaces \glspl{rnn} with a set of \glspl{mhsa} and linear layers. Like Conv-TasNet, SepFormer is a masking approach in the time domain. 
The availability of spatial information from multiple microphones allows for location-based training to resolve the permutation ambiguity issue, which further improves speaker separation results \cite{taherian2022multi}.

While most of the effective monaural speaker separation algorithms operate in the time domain, recently, \glspl{dnn} operating in the frequency domain have gained prominence by harnessing various forms of spectral information, including full-band/cross-band and sub-band/narrow-band for both single- and multi-channel speech separation. The representative model of TF-GridNet \cite{wang2022tfjournal} employs cross-band and narrow-band \gls{lstm} networks in conjunction with a cross-frame self-attention module to perform complex spectral mapping \cite{williamson2015complex,fu2017complex,tan2019learning,wang2020deep}. The most effective TF-GridNet model comprises a two-stage \gls{dnn} with a neural beamformer positioned in the intermediate stage. This model has strongly improved speech separation results in a variety of single-channel and multi-channel tasks. SpatialNet \cite{quan2023spatialnet} shares a foundational framework with TF-GridNet, but employs a combination of a Conformer narrow-band block and a convolutional-linear cross-band block. Notably, SpatialNet excludes any \gls{lstm} or \gls{rnn} layers. Furthermore, SpatialNet operates as a single-stage network and exhibits a more stable training trajectory, especially under conditions involving half-precision (16-bit) training. SpatialNet demonstrates very competitive results in multi-channel speaker separation. But its utility is primarily tailored for multi-channel scenarios, given its substantial reliance on spatial information afforded by microphone arrays. Another notable limitation of SpatialNet, in comparison to TF-GridNet, is its performance degradation with increasing sequence length, as recently reported in \cite{taherian2023multi}.

To overcome the aforementioned shortcomings and further enhance the performance of complex spectral mapping for speaker separation, we examine the underlying reasons behind the observed performance differences between TF-GridNet and SpatialNet, particularly in scenarios involving monaural separation and long utterances. We attribute the observed performance degradation of SpatialNet relative to TF-GridNet to two primary factors. First, the self-attention module within TF-GridNet operates as a global attention mechanism, whereas SpatialNet processes each frequency independently, unable to benefit from cross-frequency and hidden features. We believe that the lack of such global attention contributes to SpatialNet's diminished performance in processing long sequences. 
Second, \glspl{rnn} as exemplified by \gls{lstm} possess the capability to implicitly extract positional information \cite{rosendahl2019analysis,chen2020dual,andayani2022hybrid}. Therefore, even though neither SpatialNet nor TF-GridNet architecture explicitly incorporates \gls{pe}, the use of \glspl{rnn} captures positional cues in TF-GridNet implicitly.

In this study, we propose a new \gls{dnn} architecture, called CrossNet, for single- and multi-channel speaker separation. Building upon complex spectral mapping and the SpatialNet framework, we make the following contributions:
\begin{itemize}
    \item We present a new \gls{dnn} architecture for both single- and multi-channel speaker separation tasks. This architecture employs a global multi-head self-attention module to capture cross-frequency and cross-embedding correlations.

    \item We introduce a novel positional encoding method to CrossNet to address the out-of-distribution problem of common \gls{pe} methods.

    \item CrossNet advances the state-of-the-art speaker separation performance on multiple benchmark datasets. In addition, superior results are achieved with a reduced computational overhead
in terms of both inference and training time.

\end{itemize}

The rest of the paper is organized as follows. Section \ref{sec:problem_statement} describes the single- and multi-channel speaker separation problem in the \gls{tf} domain. The detailed description of CrossNet is given in Section \ref{sec:methodology}. Section \ref{sec:expperimental_setup} presents the experimental setup. Evaluation and comparison results are provided in Section \ref{sec:results}. Concluding remarks are given in Section \ref{sec:conclusion}.

\section{Problem statement}
\label{sec:problem_statement}

For a mixture of $C$ speakers in a noisy-reverberant environment captured by an array of $M$ microphones, the recorded mixture in the time domain $\mathbf{y}(n) \in \mathbb{R}^M$ can be modeled in terms of the direct-path signals $\mathbf{s}_c(n) \in \mathbb{R}^M$, their reverberations $\mathbf{h}_c(n) \in \mathbb{R}^M$, and reverberant background noises $\mathbf{v}(n) \in \mathbb{R}^M$

\nomenclature{$n$}{{discrete time}\nommodule{Sec \ref{sec:problem_statement}}}
\nomenclature{$C$}{{number of speakers}\nommodule{Sec \ref{sec:problem_statement}}}
\nomenclature{$c$}{{speaker subscript}\nommodule{Sec \ref{sec:problem_statement}}}
\nomenclature{$M$}{{number of microphones}\nommodule{Sec \ref{sec:problem_statement}}}
\nomenclature{$y$}{{mixture signal}\nommodule{Sec \ref{sec:problem_statement}}}
\nomenclature{$s_c$}{{direct-path signal}\nommodule{Sec \ref{sec:problem_statement}}}
\nomenclature{$h$}{{reverberation}\nommodule{Sec \ref{sec:problem_statement}}}
\nomenclature{$v$}{{reverberant background noises}\nommodule{Sec \ref{sec:problem_statement}}}

\begin{equation}
    \mathbf{y}(n) = \sum_{c=1}^C \left( \mathbf{s}_c(n)+\mathbf{h}_c(n) \right) + \mathbf{v}(n)
\end{equation} where $n$ denotes discrete time and $c$ indexes speakers. In the \gls{stft} domain, the model is expressed as:

\begin{equation}
\label{eq:tfdomain}
    \mathbf{Y}(t,f)=\sum_{c=1}^C \left( \mathbf{S}_c(t,f)  +\mathbf{H}_c(t,f) \right) + \mathbf{V}(t,f)
\end{equation} where $t$ indexes time frames and $f$ frequency bins. $\mathbf{Y}(t,f)$, $\mathbf{S}_c(t,f)$, $\mathbf{H}_c(t,f)$, and $\mathbf{V}(t,f) \in \mathbb{C}^M$ denote the complex spectrograms of the mixture, the direct-path signal and its reverberation of speaker $c$, and background noise, respectively.

The goal of complex spectral mapping based speaker separation is to train a \gls{dnn} to estimate the real and imaginary parts of the direct-path signal of each speaker $\mathbf{S}_c(t,f)$ from the mixture $\mathbf{Y}(t,f)$.
We can turn the general formulation in (\ref{eq:tfdomain}) to more specific forms by restricting certain parameters and terms. In the case of monaural, anechoic speaker separation, $C>1$, $M=1$, both $\mathbf{H}_c(t,f)$ and $\mathbf{V}(t,f)$ are absent. In reverberant speaker separation, $C>1$ and $\mathbf{V}(t,f)$, if present, represents a weak noise. In the case of noisy-reverberant speaker separation, $C>1$ and $\mathbf{V}(t,f)$ includes significant background noise.

\section{CrossNet}
\label{sec:methodology}

\begin{figure}
    \centering
    \includegraphics[width=0.7\linewidth]{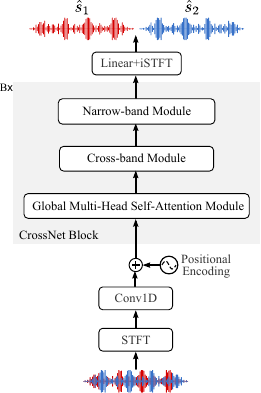}
    \caption{Diagram of the proposed CrossNet architecture, with $\hat{s}_1$ and $\hat{s}_2$ denoting separated speaker signals.}
    \label{fig:crossnet-overall}
\end{figure}

The diagram of the proposed system is provided in Fig. \ref{fig:crossnet-overall}. CrossNet comprises an encoder layer, a \gls{gmhsa} module, a cross-band module, a narrow-band module, and a decoder layer. To ensure comparable energy levels for all signals processed by CrossNet, we normalize the input signal by its variance before processing its samples. In the multi-channel setup, we normalize the signals from all microphones by the variance of the reference microphone; the same variance is applied to restore the scale of a predicted signal. Then, we apply \gls{stft} to the normalized signal and stack the \gls{ri} parts. For the multi-channel setup, we stack the \gls{ri} parts from all microphones as done in neural spectrospatial filtering \cite{tan2022neural}. The stacked \gls{ri} parts are sent to the encoder layer, which learns to extract acoustic features from the input in the \gls{stft} domain. The global multi-head self-attention module captures global correlations, while the cross-band module captures cross-band correlations. The narrow-band module focuses on capturing information at neighboring frequency bins. Finally, the output layer maps the separated features to a \gls{tf} representation, which is then converted back to the time domain using \gls{istft}.

\begin{figure*}
    \centering
    \includegraphics[width=0.8\linewidth]{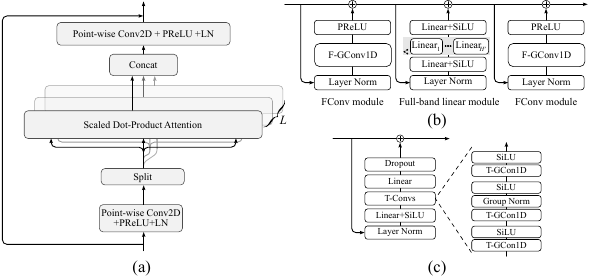}
    \caption{CrossNet building blocks. (a) Global multi-head self-attention module. (b) Cross-band module. (c) Narrow-band module. }
    \label{fig:tf-crossnet-modules}
\end{figure*}

\subsection{Encoder layer}
\label{sec:encoder}
 The encoder is a \gls{conv1d} layer with a kernel size of $k$ and a stride of 1. The encoder layer converts the input \gls{tf} domain signal from $M \times F \times T$ to $H \times F \times T$ where $H$ is the number of hidden channels. $F$ is the number of frequency bins and $T$ is number of frames.

\nomenclature{$k$}{{\gls{conv1d} kernel size}\nommodule{Encoder layer}}
\nomenclature{$H$}{{number of hidden channels}\nommodule{Encoder layer}}
\nomenclature{$T$}{{number of frames}\nommodule{Encoder layer}}
\nomenclature{$F$}{{number of frequency bins}\nommodule{Encoder layer}}

\subsection{Random chunk positional encoding}

To address the limitation of separation methods in dealing with long utterances, we introduce a \gls{pe} method, called \gls{rcpe}, to tackle the out-of-distribution problem in positional encoding approaches. \gls{rcpe} is inspired by random positional encoding recently proposed for natural language processing \cite{ruoss2023randomized}. Transformers demonstrate impressive generalization capabilities on learning tasks with a fixed context length. However, their performance degrades when tested on longer sequences than the maximum length encountered in training. This degradation is attributed to the fact that positional encoding becomes out-of-distribution for longer sequences, even for relative positional encoding \cite{ruoss2023randomized}. 
\gls{rcpe} selects a contiguous chunk of positional embedding vectors from a pre-computed \gls{pe} matrix during training. For \gls{rcpe}, we start by defining $\text{PE}$ as a combination of sine and cosine functions \cite{vaswani2017attentionisall} as 

\begin{subequations}
    \begin{align}
        \text{{PE}}(t, 2i) =   & \sin\left(\frac{{t}}{{10000^{2i/(F\times H))}}}\right) \\
        \text{{PE}}(t, 2i+1) = & \cos\left(\frac{{t}}{{10000^{2i/(F\times H)}}}\right)
    \end{align}
\end{subequations}

 When the model is in the training mode, we select a random chunk from index $\tau$ to index $\tau+T-1$, where $\tau$ is drawn randomly from $[1, T^{\text{max}}-T+1]$, with $T^{\text{max}}$ denoting the maximum desired sequence length during inference. When the model is in test or validation mode, we select the first $T$ embedding vectors. Finally, we add the selected \glspl{pe} to the input features. We obtain positional encoding vectors as

\begin{equation}
    \text{RCPE}(T) =
    \begin{cases}
        \begin{aligned}
             & \text{PE}[\tau:\tau+T-1, \dots] & \text{if training} \\
             & \text{PE}[1:T-1, \dots]   & \text{else}
        \end{aligned}
    \end{cases}
\end{equation}

This technique allows the CrossNet model to see all possible positional embedding vectors during the training stage while maintaining the relative distance between embeddings, thus improving generalization to longer sequences. Additionally, \gls{rcpe} has no learnable parameter and has a negligible computational cost.

\nomenclature{$t$}{{frame index}\nommodule{\gls{rcpe}}}
\nomenclature{$T^{\text{max}}$}{{max test frame length }\nommodule{\gls{rcpe}}}
\nomenclature{$T$}{{number of time frames in the input}\nommodule{\gls{rcpe}}}
\nomenclature{$i$}{{feature index}\nommodule{\gls{rcpe}}}
\nomenclature{$\tau$}{{random time index}\nommodule{\gls{rcpe}}}
\nomenclature{$PE$}{{positional embedding matrix}\nommodule{\gls{rcpe}}}

\subsection{Global multi-head self-attention module}
Fig. \ref{fig:tf-crossnet-modules}(a) shows the diagram of the global multi-head self-attention module. In TF-GridNet \cite{wang2022tfjournal}, the cross-frame self-attention module employs three point-wise convolution layers for frame-level feature extraction of queries, keys, and values. In contrast, we utilize a single convolution layer with $L(2E+H/L)$ output channels to extract frame-level features from \gls{tf} embeddings. Subsequently, we split the result into $L$  queries $Q^l\in\mathbb{R}^{E \times F \times T}$, keys $K^l\in\mathbb{R}^{E \times F \times T}$, and values $V^l \in\mathbb{R}^{H/L \times F \times T}$.
Here, $E$ represents the output channel dimension of the point-wise convolution and $l$ indexes the head number. This method avoids the sequential operations of the three \gls{conv1d} layers, which can be computationally expensive. Subsequently, a self-attention layer is applied to these embeddings to capture global correlations. The results of all heads are concatenated and passed to another point-wise convolution with an output dimension of $D$ followed by a \gls{prelu} activation function and \gls{lnorm}. We add this value to the input of the \gls{gmhsa} module to obtain the output of the module. 
 Note that, compared to \cite{quan2023spatialnet} where the \gls{mhsa} module acts on each frequency bin separately, we first merge all frequency features into the channel dimension and then apply \gls{mhsa}. This method allows each frame to attend to any frame of interest in all feature channels, facilitating the exploitation of long-range correlations in both frequency and hidden feature channels.

\nomenclature{$E$}{{point-wise convolution channels}\nommodule{\gls{gmhsa}}}
\nomenclature{$L$}{{number of heads}\nommodule{\gls{gmhsa}}}
\nomenclature{$l$}{{head index}\nommodule{\gls{gmhsa}}}
\nomenclature{$Q$}{{query matrix}\nommodule{\gls{gmhsa}}}
\nomenclature{$K$}{{key matrix}\nommodule{\gls{gmhsa}}}
\nomenclature{$V$}{{value matrix}\nommodule{\gls{gmhsa}}}

\subsection{Cross-band module}

To capture cross-band correlations within the input signal, we adopt the cross-band module proposed in \cite{quan2023spatialnet}. This module, illustrated in Fig. \ref{fig:tf-crossnet-modules}(b), integrates two frequency-convolutional modules and a full-band linear module. The frequency-convolutional module aims to capture correlations between neighboring frequencies. This module includes an LN layer, a grouped convolution layer along the frequency axis (F-GConv1d), and a \gls{prelu} activation function. 
In the full-band linear module, we first employ a linear layer followed by \gls{silu} activation function to reduce the number of hidden channels from $H$ to $H'$. Then, we apply a set of linear layers along the frequency axis to capture full-band features.  Each feature channel has a dedicated linear layer denoted as $\text{Linear}_i$ for $i=1, \ldots, H'$, as shown in Fig. \ref{fig:tf-crossnet-modules}(b). Note that the parameters of these layers are shared among all CrossNet blocks. Finally, the output of the module is obtained by increasing the number of channels back to $H$ using a linear layer with \gls{silu} activation and adding to the original input of this module. 

\nomenclature{$H'$}{{full-band linear module output dim.}\nommodule{Cross-band}}

\subsection{Narrow-band module}
As illustrated in Fig. \ref{fig:tf-crossnet-modules}(c), the narrow-band module is composed of a \gls{ln}, a linear layer followed by a \gls{silu}  activation, a \gls{tconv} layer, and a final linear layer. The first linear layer in this module increases the number of features in the input from $H$ to $H''$ and the last linear layer converts the feature dimension back to $H$. 

\nomenclature{$H''$}{{1$^{st}$ linear layer output dim.}\nommodule{Narrow-band}}

T-Conv is composed of three \gls{tgcon1d} layers followed by a \gls{silu} activation function. The second T-GConv1D is followed by a grouped normalization layer. The narrow-band module is a modified version of the Conformer convolutional block \cite{gulati2020conformer}. Compared to SpatialNet's narrow-band block, we remove the \gls{mhsa} module as narrow-band correlations are captured in the \gls{gmhsa} module of CrossNet.

\subsection{Output layer} 
We use a linear output layer to map the processed features from the final CrossNet block to the predicted \gls{ri} parts of each talker. Subsequently, we obtain the time-domain separated speech signals by performing the \gls{istft}. As mentioned at the beginning of Section \ref{sec:methodology}, we multiply the estimated target signals by the variance of the input mixture to ensure that their energy levels are consistent with the mixture level.

\subsection{Loss functions}
We use the \gls{si-sdr} \cite{le2019sisnr} loss function $\mathcal{L}_{\text {SI-SDR}}$  to train CrossNet on the WSJ0-2mix dataset \cite{hershey2016deep}. For training on other datasets, we employ a combination of magnitude loss $\mathcal{L}_{\text {Mag}}$ and \gls{si-sdr} loss $\mathcal{L}_{\text {SI-SDR}}$, similar to \cite{wang2022tfjournal}. We use the standard form of \gls{si-sdr} where the target signal is scaled to match the scale of the estimated signal. Also, we scale the magnitude loss by the  $L_1$ norm of the magnitude of the target signal in the \gls{stft} domain similar to \cite{pan2022hybrid}. These loss functions are defined below

\begin{subequations}
    \label{eq:loss}
    \begin{align}
        & \mathcal{L}= \mathcal{L}_{\text{Mag}} + \mathcal{L}_{\text{SI-SDR}} \\
        & \mathcal{L}_{\text{Mag}} = \frac{\left\| | \operatorname{STFT}(\hat{\boldsymbol{s}}_c) | - | \operatorname{STFT}(\boldsymbol{s}_c) | \right\|_1}{\left\| | \operatorname{STFT}(\boldsymbol{s}_c) | \right\|_1} \\
        & \mathcal{L}_{\text{SI-SDR}} = -\sum_{c=1}^C 10 \log_{10} \frac{\|\boldsymbol{s}_c\|_2^2}{\|\hat{\boldsymbol{s}}_c - \alpha_c \boldsymbol{s}_c\|_2^2} \\
        & \alpha_c = \frac{\boldsymbol{s}_c^T \hat{\boldsymbol{s}}_c}{\boldsymbol{s}_c^T \boldsymbol{s}_c}
    \end{align}
\end{subequations}

In the above equations, $\left\|\cdot\right\|_1$ is the $L_1$ norm, $|\cdot|$ is the magnitude operator, $\alpha_c$ is the scaling factor, and $(\cdot)^T$ denotes the transpose operation. We employ utterance-level \gls{pit} \cite{yu2017permutation} to resolve the permutation ambiguity problem during training.

\section{Experimental Setup}
\label{sec:expperimental_setup}

\subsection{Datasets}
We assess the efficacy of the proposed CrossNet model for speaker separation under anechoic, reverberant, and noisy-reverberant environments. we use publicly available datasets, and compare with previously published results to document the relative performance.

For single-channel speaker separation in anechoic conditions, we employ the WSJ0-2mix dataset \cite{hershey2016deep}, which is widely used for benchmarking monaural talker-independent speaker separation algorithms. The WSJ0-2mix dataset consists of 20,000 ($\sim$30.4 hours), 5,000 ($\sim$7.7 hours), and 3,000 ($\sim$4.8 hours) two-speaker mixtures for training, validation, and test sets, respectively. In WSJ0-2mix, the two utterances in each mixture are fully overlapped, and their relative energy level is sampled from the range of {[-5, 5] dB}. Speech is sampled at a rate of 8 kHz. To make a fair comparison, similar to TF-GridNet, we do not utilize any data augmentation techniques such as dynamic-mixing \cite{zeghidour2021wavesplit} or speed-perturbation \cite{subakan2021attention}.

For joint speaker separation, denoising, and dereverberation, we employ the WHAMR! dataset \cite{maciejewski2020whamr} and the single-channel SMS-WSJ dataset \cite{drude2019sms}. WHAMR! utilizes the two-speaker mixtures from WSJ0-2mix, but introduces reverberation to each clean anechoic signal and non-stationary background noises. The dataset includes 20,000 ($\sim$30.4 hours), 5,000 ($\sim$7.7 hours), and 3,000 ($\sim$4.8 hours) mixtures for training, validation, and testing, respectively.

Furthermore, for both monaural and multi-channel separation in noisy and reverberant environments, we employ the SMS-WSJ dataset \cite{drude2019sms}. This simulated two-speaker mixture dataset incorporates clean speech signals from the WSJ0 corpus and simulates a six-microphone circular array with a radius of 10 cm. \Glspl{rir} are generated using the image method \cite{allen1979image}, with T60 uniformly sampled between 0.2 s and 0.5 s. Additionally, white sensor noise is added to speech mixtures with \glspl{snr} uniformly sampled in the range of 20 dB to 30 dB. The source positions are randomly sampled within 1 m to 2 m away from the array center. The signals are sampled at a rate of {8 kHz}, and thee dataset includes a baseline \gls{asr} model built from Kaldi \cite{povey2011kaldi}.

\subsection{Network configuration}
\label{sec:network_config}

\nomenclature{$B$}{{number of CrossNet blocks}\nommodule{Fig. 1 and Sec \ref{sec:network_config}}}

For our proposed CrossNet architecture, we make use of the hyperparameters in \cite{wang2022tfjournal} and \cite{quan2023spatialnet}. We set the kernel size of encoder layer $k$, time-dimension group convolution (T-GConv1d), and frequency-dimension group convolution (F-GConv1d) to 5, 5, and 3, respectively. The number of groups for T-GConv1d, F-GConv1d, and group normalization is all set to 8. The proposed model architecture comprises $B = 12$ blocks, with hidden channel sizes set to $H = 192$, $H' = 16$, and $H'' = 384$. We employ $N=4$ self-attention heads in the \gls{gmhsa} module with an embedding dimension of $D = 64$ and $E=\left\lceil 512/F \right\rceil$, where $\left\lceil \cdot \right\rceil$ denotes ceiling operation.

To process the input data, we apply \gls{stft} using a Hanning window with frame length of 256 samples ({32 ms}) and frame shift of 128 samples ({16 ms}). The length of training utterances is fixed at 3 seconds for the WSJ0-2mix dataset and 4 seconds for the WHAMR! and SMS-WSJ datasets.

We utilize the Adam optimizer with a maximum learning rate of 0.001. We start with a cosine warm-up scheduler that increases the learning rate from $10^{-6}$ to $10^{-3}$ over the first 10 epochs. Following this, we switch to the PyTorch ReduceLROnPlateau scheduler, setting the patience to 3 epochs and the reduction factor to 0.9. We found that this learning rate scheduler is more stable and results in faster convergence than the exponential decay or ReduceLROnPlateau schedulers used in \cite{quan2023spatialnet} and \cite{wang2022tfjournal}, respectively. In our experiments, we employ the half-precision (mixed-16) training strategy to reduce the memory footprint and accelerate training. We train the model until the validation loss does not improve for 10 epochs consecutively. In each case, we use the maximum number of batches that fit into the GPU memory (NVIDIA A100 GPU with {40 GB}).

\subsection{Evaluation metrics}

We employ a set of widely used objective metrics to assess the performance of CrossNet. These metrics include: \gls{si-sdr} and its improvement (SI-SDRi) \cite{le2019sdr}, SDR and its improvement (SDRi) \cite{vincent2006performance}, narrow-band \gls{pesq} \cite{rix2001pesq}, and  \gls{estoi} \cite{jensen2016algorithmESTOI}. To compute these metrics, we utilize the TorchMetrics[audio] package \cite{detlefsen2022torchmetrics}, which offers a comprehensive set of evaluation tools specifically designed for audio tasks.

\section{Evaluation Results}
\label{sec:results}

\subsection{Ablation study on WHAMR!}

Table \ref{tab:ablation} presents an ablation study conducted on the single-channel WHAMR! dataset. Each row represents a different configuration of the model. The columns in this table provide information about the presence of \gls{rcpe}, \gls{gmhsa}, and \gls{nb-mhsa}, along with the number of trainable parameters in millions or Params (M), and the number of \glspl{gflop} per second of input audio, as well as the separation performance metrics of \gls{si-sdr}, SDR, and \gls{pesq}.  For the computation of GFLOPs, we use the official tool provided by PyTorch\footnote{{torch.utils.flop\_counter.FlopCounterMode}}.
 The absence of \gls{rcpe} and \gls{gmhsa} (Row 1) results in lower \gls{si-sdr} and \gls{pesq} scores compared to the configurations where these components are present. Note that Row 1 corresponds to the architecture of SpatialNet \cite{quan2023spatialnet}. Adding the \gls{gmhsa} module in Row 2 improves \gls{si-sdr} by {1.3 dB} and \gls{pesq} by 0.31, highlighting the important role of \gls{gmhsa}. In the third row, we include an LSTM encoder before CrossNet blocks, which performs positional encoding implicitly. The LSTM encoder comprises two \gls{blstm} layers similar to TF-GridNet's intra-frame full-band and sub-band temporal modules. Although this configuration exhibits the highest \gls{si-sdr} and \gls{pesq} values among the tested configurations, it has the largest number of parameters and the lowest computational efficiency. Including \gls{rcpe} in the fourth row improves \gls{si-sdr} by {0.3 dB} and \gls{pesq} by 0.07 compared to the second row, demonstrating the utility of the proposed positional encoding. 

 Finally, in Row 5, we remove \gls{nb-mhsa} and obtain speaker separation results with only a 0.01 \gls{pesq} reduction compared to Row 4. But the configuration with no \gls{nb-mhsa} has about 20\% fewer trainable parameters and 33\% fewer \glspl{gflop}. This shows that the narrow-band correlations are already captured in the \gls{gmhsa} module and there is little need to include both modules in the network.

 \begin{table}[H]

    \caption{Ablation study on the WHAMR! dataset }
    \centering

    \renewcommand\arraystretch{1.2}
    \setlength{\tabcolsep}{1.pt}
    \begin{tabularx}{0.95\linewidth}{ccccXccccc}
        \toprule
        Row & PE                 & \gls{gmhsa} & NB-MHSA && Params (M) & \glsfmtshort{gflop}s    & & SI-SDR  & PESQ \\ 
        \hline
        1   & \xmark             & \xmark      & \cmark  && 6.50    & 118.84 & & 10.2    & 2.54    \\
        2   & \xmark             & \cmark      & \cmark  && 8.35    & 143.51 & & 11.5    & 2.85    \\
        3   & \glsfmtshort{lstm} & \cmark      & \xmark  && 9.40    & 160.19 & & 11.9    & 2.94    \\
        4   & \glsfmtshort{rcpe} & \cmark      & \cmark  && 8.35    & 143.5  & & 11.8    & 2.92    \\
        \hline
        5   & \glsfmtshort{rcpe} & \cmark      & \xmark  && 6.57    & 96.14  & & 11.8    & 2.91    \\
        \bottomrule
    \end{tabularx}

    \label{tab:ablation}
\end{table}

\subsection{WSJ0-2mix results}

We first evaluate the performance of CrossNet for monaural anechoic speaker separation. The mixture SI-SDR is {0 dB}, and the mixture SDR is {0.2 dB}. The results are provided in Table \ref{tab:wsj0-2mix} along with 16 other baselines. 
The table includes two versions of TF-GridNet, one with 8.2M parameters and another with {14.5M} parameters. The original study \cite{wang2022tfjournal} reports the {14.5M} parameter version on the WSJ0-2mix dataset. To compare models of comparable sizes, we include the 8.2M variant as well.  The performance of this smaller TF-GridNet model is based on a model checkpoint trained by its original first author \cite{zhong_qiu_wang_2023_7565926}.
CrossNet surpasses the performance of state-of-the-art methods, including TF-GridNet ({8.2M}) \cite{wang2022tfjournal} by 0.5 dB \gls{si-sdr} and {0.6 dB} SDR. Moreover, our proposed model has around 20\% fewer trainable parameters compared to TF-GridNet and faster training and inference as presented in Section \ref{sec:flops} later.  
Furthermore, our proposed model underwent half-precision floating-point training rather than full-precision training done in TF-GridNet, effectively reducing memory requirements and expediting the training process.

\begin{table}[H]
    \caption{Speaker separation results of CrossNet and comparison methods on the WSJ0-2mix dataset }
    \begin{center}

        \renewcommand\arraystretch{1.2}
        \begin{tabularx}{0.95\linewidth}{Xccc}

            \toprule
            Method                                 & Params (M) & SI-SDRi & SDRi \\
            \midrule
            Conv-TasNet \cite{luo2019conv}          & 5.1        & 15.3 & 15.6 \\
            Deep CASA \cite{liu2019divide}          & 12.8         & 17.7    & 18.0 \\
            FurcaNeXt \cite{zhang2020furcanext}     & 51.4         & -       & 18.4 \\
            SUDO RM-RF \cite{tzinis2020sudo}        & 2.6          & 18.9    & -    \\
            DPRNN \cite{luo2020dprnn}               & 7.5          & 20.1    & 20.4 \\
            DPTNet \cite{chen2020dual}              & 2.7          & 20.2    & 20.6 \\
            DPTCN-ATPP \cite{zhu2021dptcn}          & 4.7          & 19.6    & 19.9 \\
            SepFormer \cite{subakan2021attention}   & 26.0         & 20.4    & 20.5 \\
            Sandglasset \cite{lam2021sandglasset}   & 2.3          & 20.8    & 21.0 \\
            Wavesplit \cite{zeghidour2021wavesplit} & 29.0         & 21.0    & 21.2 \\
            TFPSNet \cite{yang2022tfpsnet}          & 2.7          & 21.1    & 21.3 \\
            DPTNet \cite{chen2020dual}              & 4.0          & 21.5    & 21.7 \\
            SFSRNet \cite{rixen2022sfsrnet}         & 59.0         & 22.0    & 22.1 \\
            QDPN \cite{rixen2022qdpn}               & 200.0        & 22.1    & -    \\
            TF-GridNet$^*$ \cite{wang2022tfjournal} & 8.2          & 22.8    & 22.9 \\
            TF-GridNet \cite{wang2022tfjournal}     & 14.5         & 23.5    & 23.6 \\
            \midrule
            CrossNet                                & 6.6          & 23.2    & 23.4 \\
            \bottomrule
        \end{tabularx}
        \vspace{0.5em}

    \end{center}
    \footnotesize{$^*$checkpoint from \cite{zhong_qiu_wang_2023_7565926}}

    \label{tab:wsj0-2mix}
\end{table}
\vspace{-0.5em}

\subsection{Results on WHAMR! and single-channel SMS-WSJ datasets} 
The single-channel WHAMR! results are summarized in Table \ref{tab:whamr}. CrossNet achieves an SI-SDR of 11.8 dB, an SDR of 12.9 dB, and a PESQ of 2.91, outperforming the previous best of TF-GridNet (1-stage) \cite{wang2022tfjournal} and TF-GridNet (2-stage) \cite{wang2022tfjournal} by 0.16 and 0.12 \gls{pesq}, respectively. 
The 2-stage TF-GridNet consists of the first DNN followed by a \gls{scmfwf} and then the second DNN. This comparison is significant as CrossNet is a single-stage model, and a 2-stage model not only has more parameters but also takes more effort to train and deploy. Our advantage can be attributed to the use of more convolutional layers, which enables CrossNet to learn filtering operations.
Note that SpatialNet is not designed for single-channel separation tasks even though it can be applied to monaural separation. We include its results in Table \ref{tab:whamr} for reference purposes only. Compared to the results in Table \ref{tab:wsj0-2mix}, these results underscore the advantage of CrossNet over TF-GridNet for single-channel speaker separation in noisy-reverberant conditions.

To examine the impact of \gls{scmfwf} on model performance, we train CrossNet with a similar setup to the two-stage TF-GridNet, and 2-stage CrossNet results are included in  Table \ref{tab:whamr}. We observe a very small improvement. Thus, we conclude that two stages are not necessary and will not be further assessed for CrossNet. This observation shows that, compared to TF-GridNet where \gls{scmfwf} improves the performance, the Wiener filtering effects are already incorporated in CrossNet.

\begin{table}[H]
    \caption{Speaker separation results of CrossNet and comparison methods on the WHAMR! dataset }
    \centering

    \renewcommand\arraystretch{1.2}
    \setlength{\tabcolsep}{2.2pt}
    \begin{tabularx}{0.95\linewidth}{Xcccc}
        \toprule
        Method                                         & SI-SDR         & SDR            & PESQ           & eSTOI           \\
        \midrule
        Unprocessed                                   & -6.1           & -3.5           & 1.41           & 0.317           \\
        \hline
        Sepformer \cite{subakan2021attention}         & 7.9            & -              & -              & -               \\
        MossFormer \cite{zhao2023mossformer}          & 10.2           & -              & -              & -               \\
        SpatialNet (large) \cite{quan2023spatialnet}  & 10.2           & 11.2           & 2.54           & 0.772           \\
        TF-GridNet (1-stage) \cite{wang2022tfjournal} & 10.6           & 11.7           & 2.75           & 0.793           \\
        TF-GridNet (2-stage) \cite{wang2022tfjournal} & 11.2           & 12.3           & 2.79           & 0.808           \\
        \hline

            CrossNet (2-stage)                   &   12.0 &  13.1 &  2.91 &  0.824 \\
             CrossNet                            &  11.8 &  12.9 &  2.91 &  0.823 \\
        \bottomrule
    \end{tabularx}


    \label{tab:whamr}
\end{table}

Table \ref{tab:sms-wsh-1ch} presents evaluation and comparison results on the single-channel SMS-WSJ dataset, including \gls{asr} results in terms of \gls{wer} in percentage (\%) evaluated on the provided ASR model. CrossNet outperforms TF-GridNet (1-stage) \cite{wang2022tfjournal} by the substantial margin of 3.0 dB in \gls{si-sdr} and 0.29 in \gls{pesq}. Notably, CrossNet outperforms the two-stage TF-GridNet aside from the \gls{wer} score. 

\begin{table}[H]
    \caption{Speaker separation and ASR results on single-channel SMS-WSJ dataset }
    \begin{center}

        \renewcommand\arraystretch{1.2}
        \setlength{\tabcolsep}{1.8pt}
        \begin{tabularx}{1.0\linewidth}{Xcccccc }
            \toprule
            Method                                                       & SI-SDR         & SDR            & PESQ           & eSTOI           & WER                      \\
            \midrule
            Unprocessed                                                 & -5.5           & -0.4           & 1.50           & 0.441           & 78.40                    \\
            Oracle direct-path                                          & $\infty$       & $\infty$       & 4.50           & 1.000           & 6.16 \\
            \hline
            DPRNN-TasNet \cite{luo2020dual}                             & 6.5            & -              & 2.28           & 0.734           & 38.10                    \\
            SISO$_1$ \cite{wang2021multi}                               & 5.7            & -              & 2.4            & 0.748           & 28.70                    \\
            DNN$_1$+(FCP+DNN$_2$)$\times$2 \cite{wang2021multi}         & 12.7           & 14.1           & 3.25           & 0.899           & 12.80                    \\
            DNN$_1$+(msFCP+DNN$_2$)$\times$2 \cite{wang2021convolutive} & 13.4           & -              & 3.41           & -               & 10.90                    \\
            TF-GridNet \cite{wang2022tfjournal} (1-stage)               & 16.2           & 17.2           & 3.45           & 0.924           & 9.49                     \\
            TF-GridNet \cite{wang2022tfjournal} (2-stage)               & 18.4           & 19.6           & 3.70           & 0.952           & 7.91           \\
            \hline
             CrossNet                                          &  19.2 &  20.2 &  3.74 &  0.953 & 8.35  \\
            \bottomrule
        \end{tabularx}

    \end{center}
    \label{tab:sms-wsh-1ch}
\end{table}

\subsection{Results on multi-channel SMS-WSJ}
Table \ref{tab:sms-wsj-6ch} reports the performance of the six-channel speaker separation and ASR on the SMS-WSJ corpus, along with the oracle \gls{wer} results. The table reveals large improvements in speech quality and \gls{asr} performance thanks to speaker separation. The time-domain end-to-end models FaSNet+TAC \cite{luo2020end} and 
MC-ConvTasNet \cite{zhang2020end} show inferior performance compared to other methods, especially on the \gls{asr} task. Time-frequency methods such as MISO$_1$-BF-MISO$_3$ \cite{wang2021multi} and TFGridNet \cite{wang2022tfjournal} incorporate neural beamforming and post-processing, and demonstrate significantly better separation and \gls{asr} performances. Among the comparison methods, SpatialNet is the top performer, and it leverages an advanced full-band and sub-band combination network and extensively employs convolutional and linear layers that can act as a large filter. CrossNet surpasses SpatialNet and TFGridNet in speaker separation performance, e.g. by larger than 0.1 \gls{pesq} improvement. Since the original SpatialNet is trained using a different setup, to make a fair comparison, we also train the SpatialNet with the same setup as CrossNet, including the loss function and learning rate scheduler, and report the results in Table \ref{tab:sms-wsj-6ch}. 
Even though the use of the same training setup improves SpatialNet performance in terms of \gls{pesq}, \gls{wer}, and \gls{estoi}, it still underperformers CrossNet, e.g. by {1 dB} in SI-SDR and 5.7\% relative \gls{wer}.  CrossNet's WER of 6.30\% is remarkably close to the oracle score of 6.16\%. As CrossNet has a similar architecture to SpatialNet, the superior performance of CrossNet can be attributed to the proposed positional encoding and the \gls{gmhsa} module.

\begin{table}[H]
    \caption{Speaker separation and ASR results on the 6-channel SMS-WSJ dataset }
    \begin{center}

        \renewcommand\arraystretch{1.2}
        \setlength{\tabcolsep}{2.0pt}
        \begin{tabularx}{0.99\linewidth}{Xccccc}
            \toprule
            Method                                         & SI-SDR   & SDR      & PESQ & eSTOI & WER   \\
            \midrule
            Unprocessed                                   & -5.5     & -0.4     & 1.50 & 0.441 & 78.40 \\
            Oracle direct-path                            & $\infty$ & $\infty$ & 4.50 & 1.000 & 6.16  \\
            \hline
            FasNet+TAC \cite{luo2020end}                  & 8.6      & -        & 2.37 & 0.771 & 29.80 \\
            MC-ConvTasNet \cite{zhang2020end}             & 10.8     & -        & 2.78 & 0.844 & 23.10 \\
            MISO$_1$ \cite{wang2021multi}                 & 10.2     & -        & 3.05 & 0.859 & 14.0  \\
            LBT \cite{taherian2022multi}                  & 13.2     & 14.8     & 3.33 & 0.910 & 9.60  \\
            MISO$_1$-BF-MISO$_3$ \cite{wang2021multi}     & 15.6     & -        & 3.76 & 0.942 & 8.30  \\
            TF-GridNet (1-stage) \cite{wang2022tfjournal} & 19.9     & 21.2     & 3.89 & 0.966 & 6.92  \\
            TF-GridNet (2-stage) \cite{wang2022tfjournal} & 22.8     & 24.9     & 4.08 & 0.980 & 6.76  \\
            SpatialNet \cite{quan2023spatialnet}          & 25.1     & 27.1     & 4.08 & 0.980 & 6.70  \\
            SpatialNet$^\diamond$                         & 24.8     & 26.9     & 4.15 & 0.985 & 6.66  \\
            \hline
            CrossNet                                      & 25.8     & 27.6     & 4.20 & 0.987 & 6.30  \\

            \bottomrule
        \end{tabularx}
    \end{center}
    \footnotesize{$^\diamond$ Trained with the same setup as CrossNet}

    \label{tab:sms-wsj-6ch}
\end{table}

\subsection{Performance over different utterance lengths}
\label{sec:utt_length}
\begin{figure}
    \centering
    \includegraphics[width=0.9\linewidth]{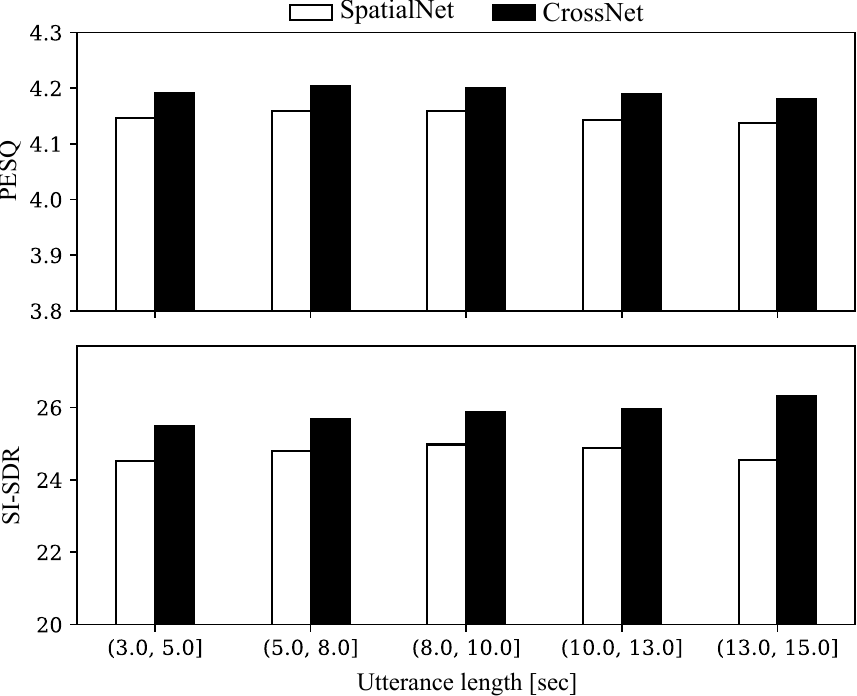}
    \caption{Effects of sequence length on the performance of CrossNet and SpatialNet. Speaker separation performance is plotted for different intervals of mixture lengths (in seconds).}
    \label{fig:pe_effect}
\end{figure}

To assess the impact of utterance length on CrossNet's performance, we plot the \gls{si-sdr} and \gls{pesq} scores across various sequence lengths on the 6-channel SMS-WSJ dataset. The results are depicted in Fig. \ref{fig:pe_effect}. CrossNet yields better performance than SpatialNet \cite{quan2023spatialnet} across all sequence lengths. The CrossNet model exhibits stable performance across different sequence lengths, whereas SpatialNet exhibits some degradation for long sequences, in line with the findings reported in \cite{taherian2023multi}, CrossNet maintains robust performance across the spectrum of sequence lengths and even produces some SI-SDR improvement for long mixtures. This performance profile highlights the contribution of the proposed positional encoding, and is important for real-world applications where the length of mixture utterances may vary significantly. 

\subsection{Computational complexity}
\label{sec:flops}
Finally, we document computational load in terms of \glspl{gflop} and the number of trainable parameters in millions (Params) of CrossNet and several other methods. The complexities are tabulated in Table \ref{tab:flops} for two sampling rates of 8 and 16 kHz. The computation of \glspl{gflop} is as outlined in \cite{quan2023spatialnet}, where \glspl{gflop} are quantified based on a four-second audio signal captured by a 6-channel microphone array speaker. The complexities of the comparison methods are obtained from \cite{quan2023spatialnet}.  As clear from the table, CrossNet exhibits much lower complexity than TF-GridNet.  Compared to SpatialNet, CrossNet has smaller GFLOPs and comparable numbers of trainable parameters.

\begin{table}
    \caption{Computational complexity and model size of the proposed model and comparison methods.}
    \begin{center}

        \renewcommand\arraystretch{1.2}
        \setlength{\tabcolsep}{1.8pt}
        \begin{tabularx}{1.0\linewidth}{Xccccc }
            \toprule
                                                 & \multicolumn{2}{c}{8 kHz} &        & \multicolumn{2}{c}{16 kHz}                  \\
            \cline{2-3} \cline{5-6}
            Model                                & \glsfmtshort{gflop}s                    & Params (M) &               & \glsfmtshort{gflop}s & Params (M) \\
            \hline
            NBDF \cite{li2019narrow}             & 19.5                     & 1.2    &                           & 38.9  & 1.2    \\
            FT-JNF \cite{tesch2022insights}      & 19.5                     & 1.2    &                           & 38.9  & 1.2    \\
            McNet \cite{yang2023mcnet}           & 29.7                     & 1.9    &                           & 59.2  & 1.9    \\
            DasFormer \cite{wang2023dasformer}   & 33.3                     & 2.2    &                           & 76.4  & 2.2    \\
            TFGridNet \cite{wang2022tfjournal}   & 348.4                    & 11.0   &                           & 695.6 & 11.2   \\
            SpatialNet \cite{quan2023spatialnet} & 119.0                    & 6.5    &                           & 237.9 & 7.3    \\
            \hline
            CrossNet                             & 96.3                     & 6.6    &                           & 191.7 & 8.2    \\
            \bottomrule
        \end{tabularx}
    \end{center}

    \label{tab:flops}

\end{table}

\section{Concluding Remarks}
\label{sec:conclusion}

We have introduced CrossNet, a novel \gls{dnn} architecture for single- and multi-channel speaker separation in noisy-reverberant environments. CrossNet includes an encoder layer, a global multi-head self-attention module, cross-band and narrow-band modules, and an output layer, to leverage both global and local information in an audio signal to enhance speaker separation and speech enhancement performance. The global multi-head self-attention module allows the model to attend to any frame of interest in all feature and frequency channels, facilitating the exploitation of long-range dependencies. We introduce a novel random chunk positional encoding technique to improve generalization to longer sequences. The cross-band module captures cross-band correlations within the input signal, while the narrow-band module focuses on capturing correlations at neighboring frequency bins. The evaluation experiments conducted on multiple open datasets demonstrate that CrossNet achieves state-of-the-art performance for single- and multi-channel speaker separation tasks. Moreover, CrossNet exhibits stable performance in separating multi-talker mixtures of variable lengths, and is computationally efficient compared to recently-established strong baselines.

\section{Acknowledgements}
We thank the members of the Perception and Neurodynamics Lab (PNL), especially Hassan Taherian and Heming Wang, for insightful discussions and their suggestions during the development of CrossNet. This work was supported in part by an National Science Foundation grant (ECCS-2125074), the Ohio Supercomputer Center, the NCSA Delta Supercomputer Center (OCI 2005572), and the Pittsburgh Supercomputer Center (NSF ACI-1928147).

\bibliographystyle{IEEEtran}
\bibliography{mybib}


\end{document}